\documentclass[ aip, PoF,
amsmath,amssymb,
reprint,%
]{revtex4-2}

\usepackage{dsfont}
\usepackage{mathtext}
\usepackage[cp1251]{inputenc}

\usepackage{mathtools}
\usepackage[usenames,dvipsnames]{xcolor}
\usepackage{amsmath}
\usepackage{amssymb,mathrsfs}
\usepackage{graphicx}
\usepackage{epstopdf}
\usepackage[colorlinks=true]{hyperref}

\usepackage{mathtext}

\usepackage{etoolbox} 
\usepackage{lipsum} 
\usepackage[capitalize]{cleveref}

\makeatletter
\appto{\appendix}{%
	\@ifstar{\def\theequation@prefix{A.}}%
	{}%
}
\makeatother

\newcommand{\om}{\omega}

\newcommand{\prt}{\partial}

\newcommand {\ox} {\overline{x}}

\newcommand{\rom}[1]{\uppercase\expandafter{\romannumeral #1\relax}}

\begin{document}

\title{Propagation of wave packets along intensive simple waves }

\author{A.~M.~Kamchatnov}
\affiliation{Institute of Spectroscopy, Russian Academy of Sciences, Troitsk, Moscow, 108840, Russia}
\affiliation{Moscow Institute of Physics and Technology, Institutsky lane 9, Dolgoprudny, Moscow region, 141700, Russia}

\author{D.~V.~Shaykin}
\affiliation{Institute of Spectroscopy, Russian Academy of Sciences, Troitsk, Moscow, 108840, Russia}
\affiliation{Moscow Institute of Physics and Technology, Institutsky lane 9, Dolgoprudny, Moscow region, 141700, Russia}

\begin{abstract}
We consider propagation of high-frequency wave packets along a smooth evolving
background flow whose evolution is described by a simple-wave type of solutions of
hydrodynamic equations. In geometrical optics approximation, the motion of the wave
packet obeys the Hamilton equations with the dispersion law playing the role of the
Hamiltonian. This Hamiltonian depends also on the amplitude of the background flow
obeying the Hopf-like equation for the simple wave. The combined system of Hamilton
and Hopf equations can be reduced to a single ordinary differential equation whose
solution determines the value of the background amplitude at the location of the
wave packet. This approach extends the results obtained in Ref.~\cite{ceh-19}
for the rarefaction background flow to arbitrary simple-wave type background flows.
The theory is illustrated by its application to waves obeying the KdV equation.
\end{abstract}




\maketitle

\section{Introduction}\label{sec1}

As is known, there is deep analogy between propagation of high-frequency wave packets and
motion of particles in classical mechanics. This analogy was discovered by Hamilton and now
it is widely used as a geometrical optics approximation to the description of wave propagation
in various non-uniform and non-stationary media (see, e.g., Refs.~\cite{synge-37,whitham-74,ko-90}
and references therein). In this approach, the wavelength of a linear wave is supposed to be much
smaller than any other characteristic size of the problem, hence such linear waves
can form a wave packet whose size is also very small and its propagation can be represented as
a motion of a point particle with the space and time dependent Hamiltonian. In particular, this
method allows one to solve various problem on propagation of high-frequency water waves
in different situations (see, e.g., \cite{mty-05,buhler-09,hoh-07}). As a result, the paths of the
wave packets along the background mean flow can be found as well as the variations of the
wave number and the amplitude.

In typical problems mentioned above, the mean flow or the medium non-homogeneity were
prescribed externally, that is, if it is characterized by a single parameter $u$, then
in the simplest one-dimensional situation the function $u=u(x,t)$ is considered as known.
Linear waves propagating along such a background flow have a dispersion law
$\om=\om(u,k,x,t)$ which relates the frequency $\om$ of the linear wave with its wave
number $k$ in vicinity of the point $x$ at the moment of time $t$.
However, instead of external prescription of the background flow $u(x,t)$, one may suppose
that the initial distribution $u(x,t=0)$ is given, so that the background evolves dynamically
and we want to consider the propagation of high-frequency wave packets along such a non-stationary
background flow. Generally speaking, this is a quite difficult problem, but it has recently
been noticed in Ref.~\cite{ceh-19} that if one confines oneself to the simple-wave type
of the background evolution, then the problem greatly simplifies and becomes tractable
analytically. In particular, the propagation of wave packets along rarefaction wave
was studied in Ref.~\cite{ceh-19} and two different possible regimes were found---transmission
of the packet through the finite rarefaction wave or trapping of the packet inside it.
In this paper, we extend the approach of Ref.~\cite{ceh-19} to arbitrary smooth profile of
the simple-wave type. To this end, we use the consistency condition
that the wave packet propagates with the group velocity along the simple-wave solution
of the dynamical dispersionless equation \cite{kamch-19a,ik-19}. Recently, such a generalization
permitted us to develop a new approach to derivation of the number of solitons produced by an
intensive initial pulse at asymptotically large time \cite{kamch-20a,ik-20,kamch-21}. In these
papers, the wave packet was identified with the small amplitude edge of a dispersive shock wave
and here we generalize this approach to an arbitrary localized wave packet injected into the
evolving simple wave profile. The general theory is illustrated by its application to the
weakly dispersive and weakly nonlinear situations described by the KdV equation. The analytical
solution agrees very well with exact numerical solutions.

\section{Propagation of wave packets along simple waves}

We assume that the dispersion law of linear waves propagating along the background flow
described by a single variable $u$ is given by the function $\om=\om(u,k)$. The
background flow is supposed to be a simple wave, so any physical variable can be
expressed in terms of $u$ and the propagation is unidirectional with some velocity
$V_0(u)$ which is equal to the long wavelength limit of the phase velocity of
linear waves (for definiteness, we consider a right-propagating wave)
\begin{equation}\label{t5-56.2}
  V_0(u)=\lim_{k\to0}\frac{\om(u,k)}k.
\end{equation}
Thus, we arrive at the Hopf equation
\begin{equation}\label{t5-56.1}
  u_t+V_0(u)u_x=0,
\end{equation}
which governs evolution of the smooth background flow.

A typical wavelength $\sim2\pi/k$ of the wave packet propagating along the background
flow is much smaller than the characteristic size $l$ of variation of $u(x,t)$.
We assume that the size of the wave packet made from these linear waves with
wave numbers close to $k$ is also much smaller than $l$. Therefore, in the geometric
optics approximation, we can define the ``mean'' coordinate $x(t)$ of the packet
and the ``mean'' or ``carrier'' wave number $k(t)$ around which the spectrum of
the packet is concentrated. Then, according to Hamilton's optical-mechanical
analogy (see, e.g. \cite{lanczos}) $x(t)$ and $k(t)$ obey the Hamilton equations
\begin{equation}\label{t5-56.3}
  \frac{dx}{dt}=\frac{\prt\om(u,k)}{\prt k},\qquad \frac{dk}{dt}=-\frac{\prt\om(u,k)}{\prt x}.
\end{equation}
In our case the dependence of the Hamiltonian $\om$ on $x$ and $t$ is determined by
the solution $u=u(x,t)$ of the equation (\ref{t5-56.1}) for the background flow.
This leads immediately to an important consequence: from
(\ref{t5-56.1}) and (\ref{t5-56.3}) we get at once
\begin{equation}\nonumber
\begin{split}
  \frac{dk}{dt}& =-\frac{\prt\om}{\prt u}\frac{\prt u}{\prt x},\\
  \frac{du}{dt}& =\frac{\prt u}{\prt x}\frac{dx}{dt}+\frac{\prt u}{\prt t}=
  -\left(V_0-\frac{\prt\om}{\prt k}\right)\frac{\prt u}{\prt x},
  \end{split}
\end{equation}
and the ratio of these expressions yields
\begin{equation}\label{t5-56.4}
  \frac{dk}{du}=\frac{\prt\om/\prt u}{V_0-\prt\om/\prt k}.
\end{equation}
This equation was derived in Ref.~\cite{el-05} as a consequence of the Whitham modulation
equations at the small-amplitude edge of dispersive shock waves. Here the right-hand side
only depends on $u$ and $k$, so the solution of this equation can be written as
\begin{equation}\label{t5-56.5}
  k=k(u,q),
\end{equation}
where $q$ is the integration constant which can be found, for example, from the condition
that the wave packet enters into the region of a non-uniform flow with the wave number $k_0$.

\begin{figure}[t]
\begin{center}
\includegraphics[width=8cm]{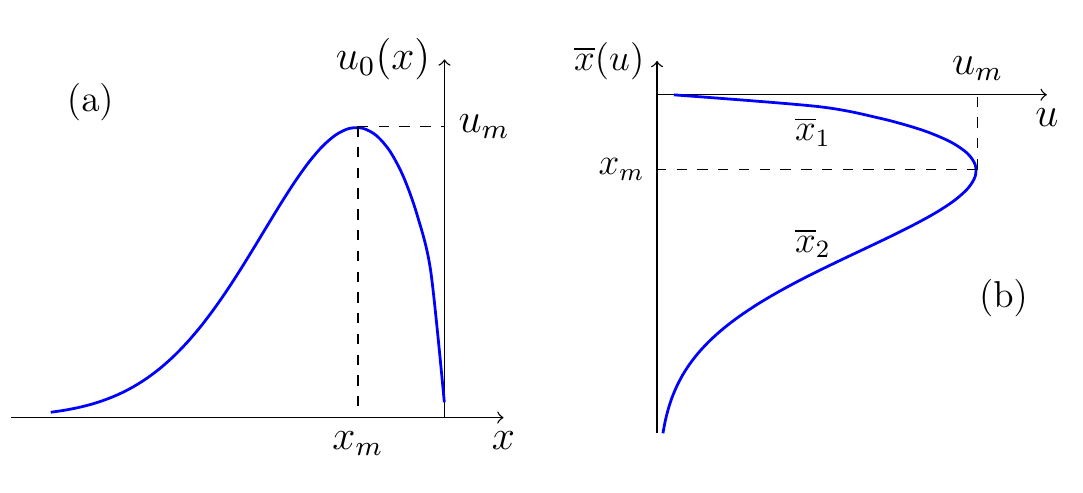}
\caption{(a) The initial profile $u=u_0(x)$ localized in the region $x\leq0$.
(b) Two branches of the inverse function $\ox_1(u)$ and $\ox_2(u)$.
}
\label{fig1}
\end{center}
\end{figure}

Now we take into account that the solution of the Hopf equation (\ref{t5-56.1})
is given by the expression (see, e.g., Ref.~\cite{whitham-74})
\begin{equation}\label{t5-57.1}
  x-V_0(u)t=\ox(u),
\end{equation}
where $\ox(u)$ denotes the function inverse to the initial distribution $u=u_0(x)$.
If the initial pulse has a form of a hump on the constant background (see., e.g.,
Fig.~\ref{fig1}(a)), then the inverse function has two branches shown in
Fig.~\ref{fig1}(b), and Eq.~(\ref{t5-57.1}) defines the solution $u=u(x,t)$ for each
branch in an implicit form. In a similar way we can consider a ``negative'' background
pulse with $u_0(x)\leq0$. As soon as we find the function $u=u(x,t)$ as a solution of 
Eq.~(\ref{t5-57.1}) and the function $k=k(u)$ as a solution (\ref{t5-56.5}) of
 Eq.~(\ref{t5-56.4}), we can find the law of motion of the
wave packet in the following way \cite{kamch-19a,ik-19}. The first Hamilton equation
(\ref{t5-56.3}) says that the wave packet moves with the group velocity
$v_g(u,k)=\prt\om(u,k)/\prt k$ and in time interval $dt$ it passes the distance $dx=v_gdt$
along the ``surface'' $u=u(x,t)$ of the smooth solution of Eq.~(\ref{t5-57.1}).
We assume that the packet's path is determined by parametric formulas
$t=t(u)$, $x=x(u)$ and find the derivative of Eq.~(\ref{t5-57.1}) with respect to $u$.
Then elimination of $dx/du=(dx/dt)(dt/du)=v_g(dt/du)$ from the expression for
this derivative yields a linear differential equation
\begin{equation}\label{t5-57.2}
  [v_g(u,k(u))-V_0(u)]\frac{dt}{du}-V_0'(u)t=\ox'(u).
\end{equation}
for the function $t(u)$ which can be easily solved with the initial condition $t=0$
at some initial value of $u$ corresponding to the initial location of the wave packet. 
Substitution of $t(u)$ into Eq.~(\ref{t5-57.1})
gives $x=x(u)$ and as a result we obtain the law of motion $x=x(t)$ in a parametric form.
If the function $\ox(u)$ is two-valued, then the solution $t=t_1(m)$ found in this way is
correct up to the moment $t_1(u_m)$ when the packet reaches the maximal (or minimal) value
of the background distribution $u=u_m$ (according to the solution (\ref{t5-57.1}) $u_m$ does
not depend on time). After that the packet moves along the second branch of the smooth
solution and Eq.~(\ref{t5-57.2}) with $\ox=\ox_2(u)$ must be solved with the initial
condition $t=t_1(u_m)$ at $u=u_m$ which gives $t=t_2(u)$. Thus, we obtain the function
$t=t(u)$ for the entire motion of the packet propagation along the evolving pulse $u=u(x,t)$.
In the next Section we shall illustrate this method by its application to two typical
problems.

\section{Example: Korteweg-de Vries equation}

To illustrate the formulated above approach, we shall apply it to waves whose
propagation obeys the Korteweg-de Vries (KdV) equation
\begin{equation}\label{t5-57.3}
  u_t+6uu_x+u_{xxx}=0.
\end{equation}
It appears in a number of physical applications for description of weakly nonlinear
and weakly dispersive unidirectional wave propagation. It was first derived \cite{kdv}
for explanation of soliton propagation on shallow water and in this paper we
shall assume this physical application of the theory.

If we neglect the last dispersive term, it reduces to the Hopf equation (see (\ref{t5-56.1}))
\begin{equation}\label{t5-57.3b}
  u_t+6uu_x=0
\end{equation}
with $V_0(u)=6u$. It describes large scale evolution of smooth waves with a characteristic
dimension $l$. We are interested in propagation along such an evolving wave of a wave
packet with the carrier wave number $k\ll 1/l$. Within such a packet the background
amplitude can be assumed constant and the linearized KdV equation
\begin{equation}\nonumber
  u'_t+6uu'_x+u'_{xxx}=0
\end{equation}
yields for harmonic linear waves $u'\propto\exp[i(kx-\om t)]$ the dispersion law
\begin{equation}\label{t5-57.4}
  \om(u,k)=6uk-k^3.
\end{equation}
Hence, the wave packet moves with the velocity
\begin{equation}\label{t5-57.4b}
  v_g(u,k)=\frac{\prt\om}{\prt k}=6u-3k^2.
\end{equation}

At first, we shall briefly consider the propagation of such a wave packet along a rarefaction
wave. This problem was discussed in much detail in Ref.~\cite{ceh-19}. We shall show
that the Hamiltonian approach permits one to simplify the solution and to
obtain some additional results which will help us to compare them with more general
situations considered in the subsequent Subsections.

\subsection{Propagation of a wave packet along a rarefaction wave}

\begin{figure}[t]
\begin{center}
\includegraphics[width=7cm]{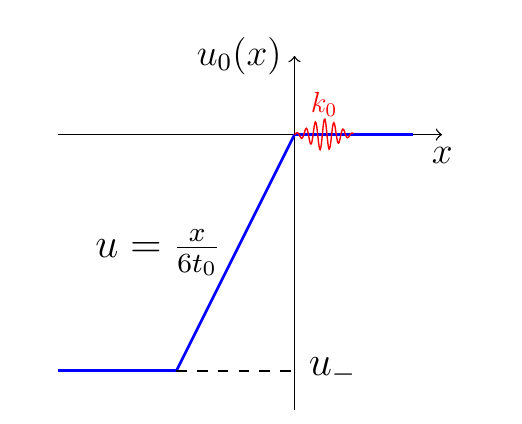}
\caption{Thick (blue) line shows the initial profile (\ref{t5-57.5}) evolving into
a rarefaction wave. The wavy (red) line symbolizes a short-wavelength wave
packet with the carrier wave number $k_0$ entering into the region of the
large scale smooth wave from the right.
}
\label{fig2}
\end{center}
\end{figure}

Let the initial distribution have the form
\begin{equation}\label{t5-57.5}
  u_0(x)=\left\{
  \begin{array}{ll}
  0,\qquad &x>0,\\
  x/(6t_0),\qquad &6u_-t_0\leq x\leq0,\\
  u_-,\qquad &x<6u_-t_0,\\
  \end{array}
  \right. ,
\end{equation}
where $u_-<0$ (see Fig.~\ref{fig2}). Then the solution of Eq.~(\ref{t5-57.3b})
is a rarefaction wave
\begin{equation}\label{t5-57.6}
  u(x,t)=\frac{x}{6(t+t_0)}\quad \text{for}\quad 6u_-(t+t_0)\leq x\leq 0,
\end{equation}
which connects the plateau regions $u=0$ for $x>0$ and $u=u_-$ for $x<6u_-(t+t_0)$.
The left edge of the rarefaction wave propagates along the constant background
$u=u_-$ to the left with the ``sound velocity'' $6u_-$.

Now let a high-frequency wave packet with the wave number $k_0$ enter into the
rarefaction wave region from the right at the moment $t=0$, so that its initial
coordinate is $x(0)=0$ and the initial velocity equals to $v_g(0)=-3k_0^2$.
The wave number $k$ varies during the propagation and we find the law of this
variation by solving Eq.~(\ref{t5-56.4}), which in our case takes the form
\begin{equation}\label{t5-57.7}
  \frac{dk}{du}=\frac2k,
\end{equation}
with the initial condition $k(0)=k_0$ to obtain
\begin{equation}\label{t5-57.8}
  k(u)=\sqrt{4u+k_0^2}.
\end{equation}

The function inverse to the initial distribution (\ref{t5-57.5}) is given by
$\ox(u)=6t_0u$, so Eq.~(\ref{t5-57.2}) transforms to
\begin{equation}\label{t5-57.9}
  2\left(u+\frac{k_0^2}4\right)\frac{dt}{du}=-(t+t_0)
\end{equation}
and we easily find its solution with the initial condition $t(0)=0$:
\begin{equation}\label{t5-57.10}
  t(u)=\frac{k_0t_0}{\sqrt{4u+k_0^2}}-t_0.
\end{equation}
Consequently, along the packet's path the background amplitude changes with time as
\begin{equation}\label{t5-58.11}
  u(t)=-\frac{k_0^2}4\cdot\frac{t(t+2t_0)}{(t+t_0)^2}
\end{equation}
and its substitution into Eq.~(\ref{t5-57.1}) yields the explicit formula for the path:
\begin{equation}\label{t5-58.12}
  x(t)=-\frac{3k_0^2}2\cdot\frac{t(t+2t_0)}{t+t_0}.
\end{equation}

If $|u_-|$ is large enough ($|u_-|>k_0^2/4$), then in the limit $t\to\infty$ the packet
moves with constant velocity
\begin{equation}\label{t5-58.13}
  v_g(\infty)=-\frac{3k_0^2}2
\end{equation}
smaller in the absolute value than the velocity $6|u_-|$ of the left edge of the
rarefaction wave, hence the packet remains forever inside the rarefaction wave
region. The wave number (\ref{t5-57.8}) depends on time as
\begin{equation}\label{t5-58.13b}
  k(t)=\frac{k_0t_0}{t+t_0},
\end{equation}
and $k\to0$ as $t\to\infty$. This means that the packet disperses with time
and we go beyond applicability of the geometric optics Hamiltonian approximation.

If $|u_-|<k_0^2/4$, then the packet passes through the rarefaction wave region
and goes out from it with the wave number
\begin{equation}\label{t5-58.14}
  k_-=\sqrt{k_0^2+4u_-}<k_0
\end{equation}
at the moment of time
\begin{equation}\label{t5-58.15}
  t_-=\left(\frac{k_0}{k_-}-1\right)t_0.
\end{equation}
The exit coordinate of the packet on the plateau with $u=u_-$ is given by
\begin{equation}\label{t5-58.16}
  x(t_-)=\frac{6k_0u_-}{k_-}\,t_0.
\end{equation}
If there were no the rarefaction wave, the packet would move to the point
with the coordinate $-3k_0^2t_-$ shifted with respect to (\ref{t5-58.16}) by
the distance
\begin{equation}\label{t5-58.17}
  \Delta x=-\frac{3k_0(k_0-k_-)^2}{2k_-}\,t_0,
\end{equation}
which can be called the ``phase shift'' caused by interaction of the packet
with the rarefaction wave.

These analytical results agree very well with the results of numerical solution
of the KdV equation and some details can be found in Ref.~\cite{ceh-19}.

\subsection{Propagation of a wave packet along a negative pulse}

\begin{figure}[t]
\begin{center}
\includegraphics[width=7cm]{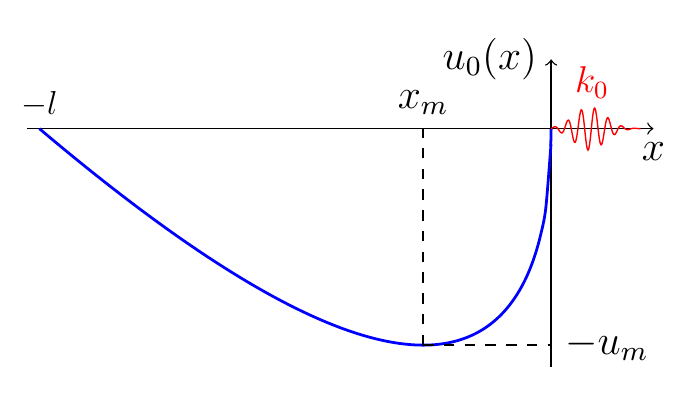}
\caption{Thick (blue) line shows the initial negative profile (\ref{t5-58.1}).
The wavy (red) line symbolizes a short wavelength wave
packet with the carrier wave number $k_0$ entering into the region of the
large scale smooth wave from the right.
}
\label{fig3}
\end{center}
\end{figure}

\begin{figure}[t]
\begin{center}
\begin{picture}(130.,120.)
\put(-60,0){\includegraphics[width = 3.95cm]{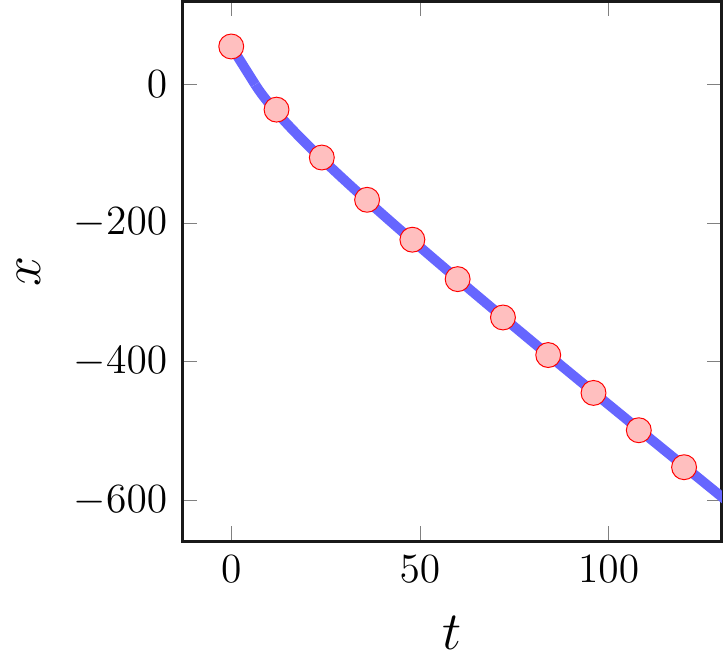}}
\put(65,0){\includegraphics[width = 4cm]{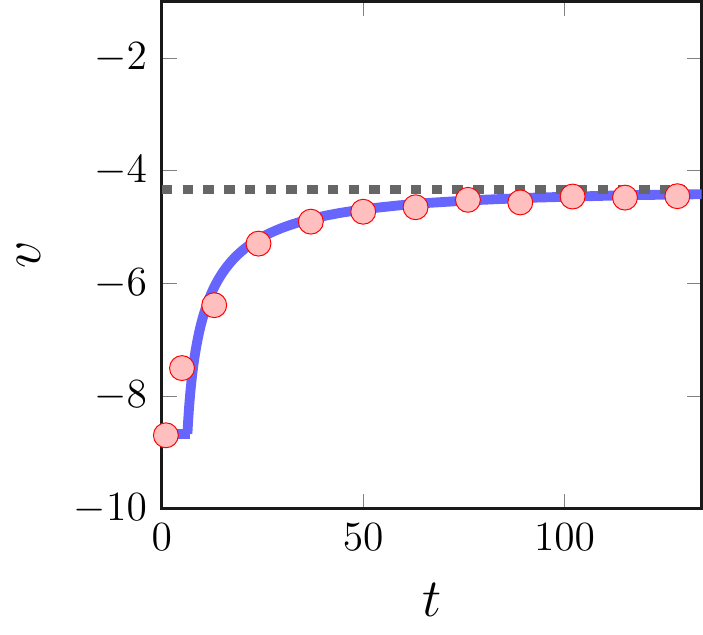}}
\put(5,110){(a)}
\put(125,110){(b)}
\end{picture}
\caption{(a) The path $x(t)$ of the wave packet with $k_0=1.7$ along the
negative pulse (\ref{t5-58.3}) with $u_m=1$ and $l=800$. This path is given in a
parametric form by
the formulas (\ref{t5-59.6}), (\ref{t5-59.6a}) and shown by a blue solid line.
Red points correspond to the numerical solution of the KdV equation.
(b) The corresponding velocity $v=dx/dt$. It tends to the limiting value
(\ref{t5-59.6b}) as $t\to\infty$. The blue solid line corresponds to the analytical
formulas and the red points to the numerical solution of the KdV equation.
}
\label{fig4}
\end{center}
\end{figure}

Now we shall turn to the motion of a high-frequency wave packet along a more general
simple-wave profile which occupies a finite region in the space. We shall start with
a ``negative'' pulse for which $u_0(x)<0$ and for definiteness we shall take the
initial profile in the form (see Fig.~\ref{fig3})
\begin{equation}\label{t5-58.1}
  u_0(x)=-4u_m\left(\frac{x}l+\sqrt{-\frac{x}l}\right),\qquad -l\leq x\leq 0,
\end{equation}
which admits a simple enough explicit analytical solution of the problem.
In this case we have two branches of the inverse function
\begin{equation}\label{t5-58.2}
\begin{split}
  &\ox_1(u)=\frac{l}4\left(-\frac{u}{u_m}-2+2\sqrt{1+\frac{u}{u_m}}\right),\\
  &\ox_2(u)=\frac{l}4\left(-\frac{u}{u_m}-2-2\sqrt{1+\frac{u}{u_m}}\right),
  \end{split}
\end{equation}
and for both branches the solution of the equation $x-6ut=\ox(u)$ with respect
to $u(x,t)$ yields the same expression
\begin{equation}\label{t5-58.3}
\begin{split}
  u(x,t)=&\frac{4u_m}{(l-24u_mt)^2}\Big[12u_mt(l+2x)-lx\\
  &-l\sqrt{24u_mt(6u_mt+x)-lx}\Big].
  \end{split}
\end{equation}
This dispersionless solution is single-valued up to the wave breaking moment $t_b$
determined by the condition $\prt u/\prt x\to\infty$ as $u\to0$ which gives
\begin{equation}\label{t5-58.3b}
  t_b=\frac{l}{12u_m}.
\end{equation}
As before, we assume that the wave packet with the wave number $k_0$ enters
into the pulse's region at the point $x=0$ at the moment of time $t=0$, and we are
interested in its motion along the evolving profile (\ref{t5-58.3}).

\begin{figure}[t]
\begin{center}
\begin{picture}(130.,120.)
\put(-59,2){\includegraphics[width = 4.31cm]{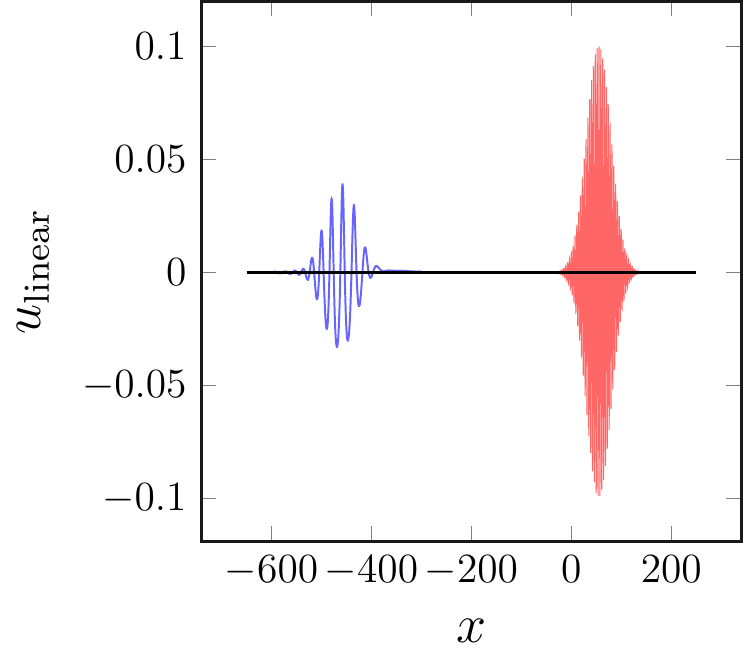}}
\put(67,0){\includegraphics[width = 4cm]{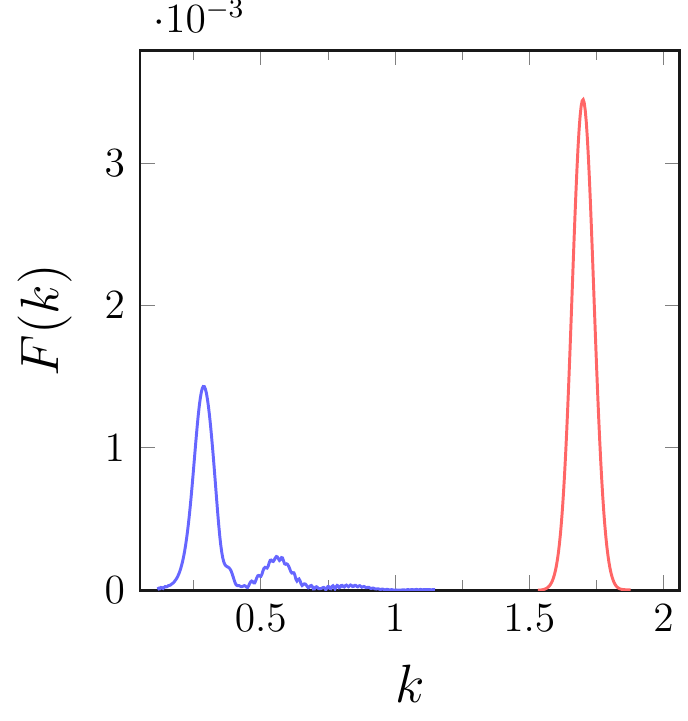}}
\put(15,115){(a)}
\put(130,115){(b)}
\end{picture}
\caption{(a) The initial profile of the wave packet at $0<x<100$ (red) and its profile
for a large value of time at $-600<x<-400$ (blue), when it is practically blocked inside the
smooth pulse. The wavelength considerably increases with time.
(b) The initial spectrum of the wave packet centered around $k_0=1.7$ (red) and its spectrum
for a large value of time $t=100$ (blue) when the carrier wave number is close to zero 
according the formula (\ref{t5-57.8}) with $u\to-k_0^2/4$. Since the packet's amplitude is not
infinitely small, the propagation of the packet is accompanied by generation of higher
harmonics with corresponding peaks in the spectrum. The peak at the local carrier wave number
$k=0.28$ and two additional peaks at $k=0.56$ and $k=0.84$ are clearly seen
in our numerical solution of the KdV equation.
}
\label{fig5}
\end{center}
\end{figure}

\begin{figure}[t]
\begin{center}
\includegraphics[width=5cm]{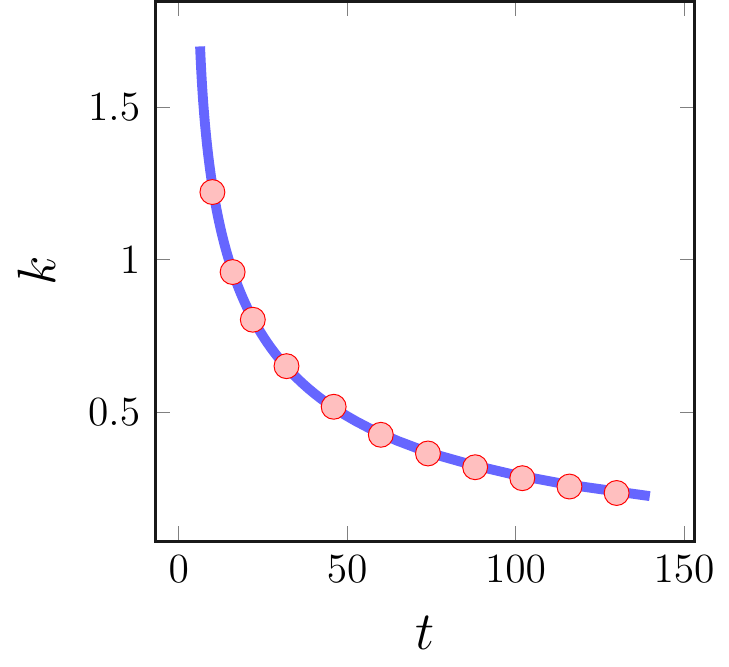}
\caption{The dependence of the carrier wave number on time for $k_0=1.7$.
The solid blue line correspond to the analytical theory and red points to
the numerical solution of the KdV equation.
}
\label{fig6}
\end{center}
\end{figure}

Equation (\ref{t5-57.7}) does not depend on the form of the profile, so the dependence
of the wave number $k(u)$ on the background amplitude is still given by Eq.~(\ref{t5-57.8}).
However, the right-hand side of Eq.~(\ref{t5-57.2}) equals now to one of the
following expressions
\begin{equation}\label{t5-59.4}
\begin{split}
  &\ox'_1(u)=-\frac{l}{4u_m}\left(1-\frac1{\sqrt{1+u/u_m}}\right),\\
  &\ox'_2(u)=-\frac{l}{4u_m}\left(1+\frac1{\sqrt{1+u/u_m}}\right),
  \end{split}
\end{equation}
depending on along which branch the packet moves. Solution of the linear equation
\begin{equation}\label{t5-59.5}
  12\left(u+\frac{k_0^2}4\right)\frac{dt}{du}+6t=-\ox'(u)
\end{equation}
with $\ox'(u)=\ox_1'(u)$ and the initial condition $t(0)=0$ gives for the motion along the
first branch the expression
\begin{equation}\label{t5-59.6}
\begin{split}
  t_1(u)&=\frac{l}{48u_m\sqrt{u+k_0^2/4}}\times\\
  &\times\int_0^u\left(1-\frac1{\sqrt{1+u/u_m}}\right)\frac{du}{\sqrt{u+k_0^2/4}}\\
  &=\frac{l}{24u_m\sqrt{k_0^2+4u}}\Bigg[\sqrt{k_0^2+4u}-k_0+\\
  &+2\sqrt{u_m}\ln\frac{k_0+2\sqrt{u_m}}{\sqrt{k_0^2+4u}+2\sqrt{u_m+u}}\Bigg].
  \end{split}
\end{equation}

If $k_0<2\sqrt{u_m}$, then $t_1\to\infty$ as $u\to-k_0^2/4$, consequently $u$ does not
reach the minimal value $u_m$ and the packet moves all the time along the first branch
corresponding to the right side of the smooth distribution point according
to the law
\begin{equation}\label{t5-59.6a}
  x(u)=6ut_1(u)+\ox_1(u).
\end{equation}
Formulas (\ref{t5-59.6}), (\ref{t5-59.6a}) determine the packet's path $x=x(t)$ in a
parametric form. In Fig.~\ref{fig4}(a) we compare the analytical theory with the
numerical solution of the KdV equation. We take the initial pulse in the form
(\ref{t5-58.1}) with $u_m=1$ and $l=800$. The initial width
of the packet is about $\Delta x\sim 50$, so we choose the initial coordinate of its
center at $x_0=55$ and the carrier wave number equal to $k_0=1.7$. The group velocity
tends in this limit to the value
\begin{equation}\label{t5-59.6b}
  v_g\approx-\frac32k_0^2\quad\text{as}\quad t\to\infty
\end{equation}
in agreement with the numerical solution (see Fig.~\ref{fig4}(b)).

In Fig.~\ref{fig5}(a) we show the packet's profile at two moments of time --- at the
initial one and at large time close to the asymptotic regime (\ref{t5-59.6b}) when the
packet is practically blocked within the smooth profile and moves with the velocity
(\ref{t5-59.6b}). To find these profiles, we solve numerically the KdV equation for
two types of the initial conditions---with and without the initial wave packet contribution.
Then subtraction of the second profile from the first one yields the packet's
amplitude $u_{\mathrm{linear}}(x,t)$ without contribution of the smooth pulse.
We see that the wavelength of
the carrier wave increases considerably in the asymptotic regime. This qualitative
observation is confirmed quantitatively by the plots of the spectrum
\begin{equation}\label{spectrum}
  F(k)=\int_{-\infty}^{\infty}u_{\mathrm{linear}}(x,t)e^{-ikx}dx
\end{equation}
shown in Fig.~\ref{fig5}(b). It is worth noticing that since in our numerical simulations
the packet's amplitude is not infinitely small, the propagation of the packet is 
accompanied by generation of higher harmonics. As a result, the spectrum acquires additional 
peaks in the wavenumber distribution and two such peaks are clearly seen in Fig.~\ref{fig5}(b).

Formula (\ref{t5-59.6a}) together with Eq.~(\ref{t5-57.8}) yields the dependence $k=k(x)$
of the carrier wave number along the packet's path. This wave number vanishes in
the asymptotic limit $u\to-k_0^2/4$ and its dependence on time is given in a parametric form
by the formulas (\ref{t5-57.8}) and (\ref{t5-59.6}). In Fig.~\ref{fig6} we compare the
analytical theory with the results extracted from the numerical solution of the KdV equation
and find very good agreement. 

The asymptotic behavior at $t\to\infty$ can be interpreted in the following way. The first
branch of the dispersionless solution (\ref{t5-58.3}) not too close to the minimum point
approaches in the limit $t\to\infty$ to the rarefaction wave solution $u(x,t)\approx x/(6t)$,
so that for small $k_0\ll\sqrt{u_m}$ the packet's motion practically coincides with the motion
considered in the preceding Subsection. The asymptotic law of motion is obtained from
Eq.~(\ref{t5-58.12}) with $t_0=0$:
\begin{equation}\label{t5-59.9}
  x(t)\approx-\frac{3u_0^2}2t,
\end{equation}
that is the wave packet moves asymptotically with the velocity (\ref{t5-59.6b}).
However, one has to keep in mind that after the wave breaking moment (\ref{t5-58.3b})
the dispersive shock wave appears at the left edge of the pulse (\ref{t5-58.3}),
and the right edge of the shock propagates according to the law (see Ref.~\cite{ikp-19})
\begin{equation}\label{t5-60.10}
  x_R^{\text{DSW}}=-\frac{3A^{2/3}}{2^{1/3}}t^{1/3},\qquad A=\int\sqrt{u_0(x)}\,dx.
\end{equation}
Consequently, after the moment $t\approx2A/k_0^3$ the packet enters into the dispersive shock
region and propagates along its averaged profile rather than along the dispersionless
solution (\ref{t5-58.3}). Thus, the above theory is applicable for the time $t<2A/k_0^3$
which may be very large for long enough initial pulse length $l$.

\begin{figure}[t]
\begin{center}
\begin{picture}(130.,120.)
\put(-55,0){\includegraphics[width = 4cm]{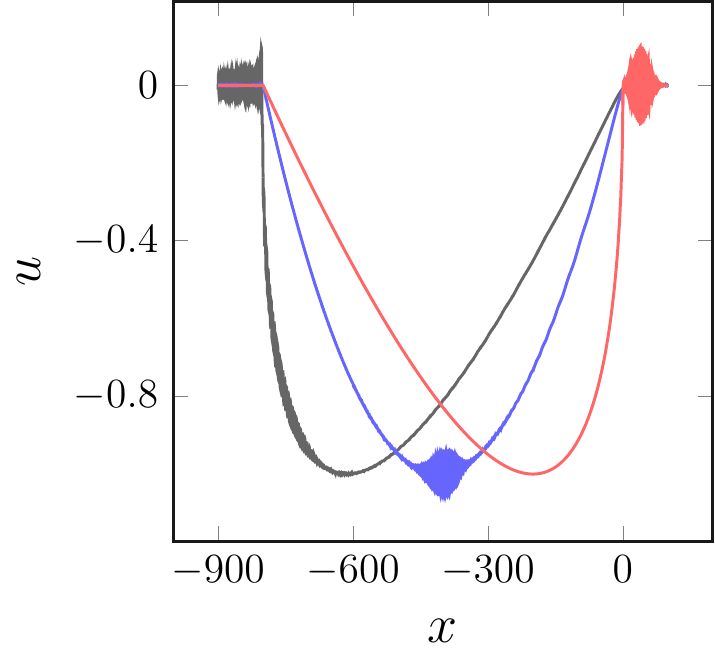}}
\put(65,-2){\includegraphics[width = 4.2cm]{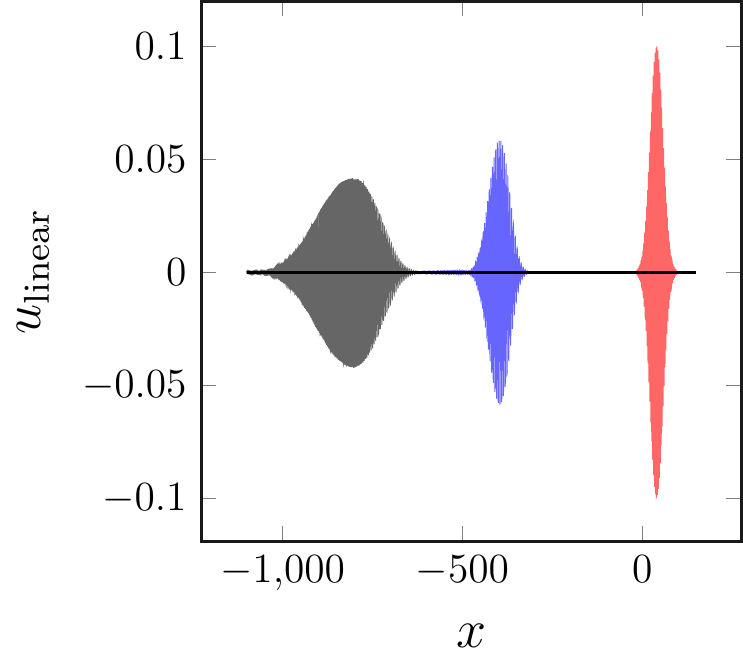}}
\put(10,110){(a)}
\put(130,110){(b)}
\end{picture}
\caption{(a) The total numerical profiles $u(x,t)$ obtained by numerical solution of
the KdV equation for $u_m=1$, $k_0=2.4$ at the moments of time $t=0$ (red), $t=35$ (blue),
$t=69$ (black). (b) The profiles of the linear wave packets obtained by subtraction of
the smooth evolution of the background pulse at the same moments of time.
}
\label{fig7}
\end{center}
\end{figure}

\begin{figure}[t]
\begin{center}
\includegraphics[width =5cm]{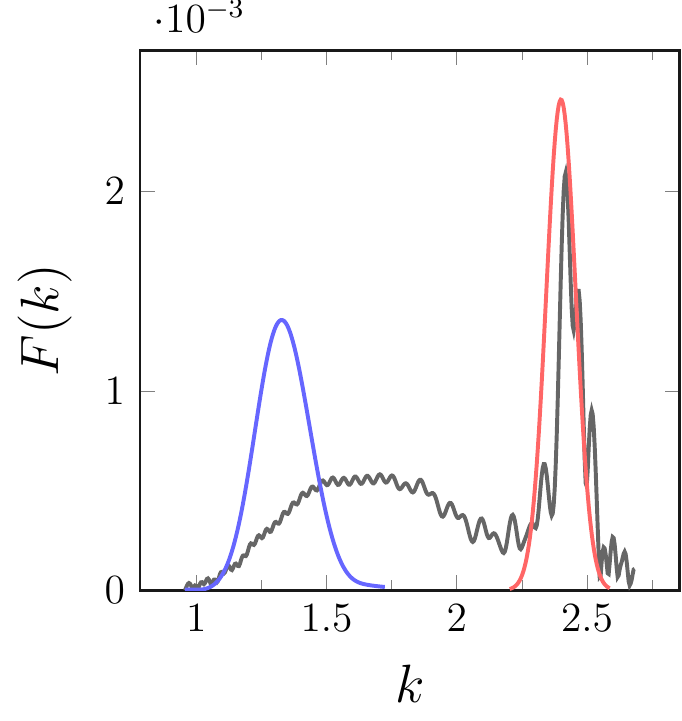}
\caption{
The spectrum of the wave packet at the same moments of time which are indicated in
Fig.~\ref{fig7}. At the moment of time $t=69$ the packet is located partly outside the
smooth pulse and partly inside it what causes two humps in the distribution. The narrow
peak in the spectral distribution corresponds to restoration of the initial carrier wave
number during the exit of the packet from the smooth pulse region.
}
\label{fig8}
\end{center}
\end{figure}

\begin{figure}[t]
\begin{center}
\includegraphics[width=5cm]{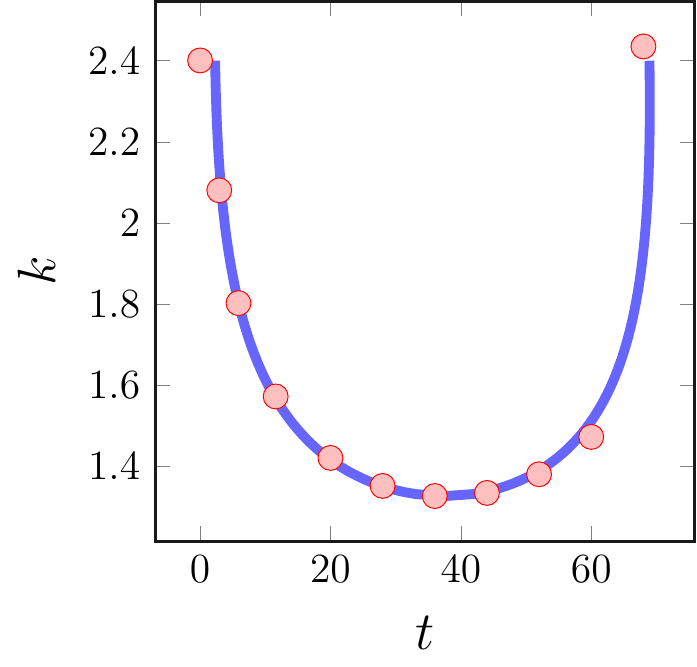}
\caption{The dependence of the carrier wave number on time for
negative pulse and the initial wave number $k_0=2.4$.
}
\label{fig9}
\end{center}
\end{figure}

If $k_0>2\sqrt{u_m}$, then the packet reaches the minimal point $u_m$ of the background
profile at the moment
\begin{equation}\label{t5-60.11}
\begin{split}
  &t_m=t_1(-u_m)=\frac{l}{24u_m\sqrt{k_0^2-4u_m}}\times\\
  &\times\left[\sqrt{k_0^2-4u_m}-k_0+\sqrt{u_m}
  \ln\frac{k_0+2\sqrt{u_m}}{k_0-2\sqrt{u_m}}\right].
  \end{split}
\end{equation}
After that it starts its motion along the second branch for which we easily obtain
\begin{equation}\label{t5-60.12}
\begin{split}
  t_2(u)=&\frac{l}{24u_m\sqrt{k_0^2+4u}}\Bigg[\sqrt{k_0^2+4u}-k_0+\\
 & +2\sqrt{u_m}\ln\frac{\sqrt{k_0^2+4u}+2\sqrt{u+u_m}}{k_0-2\sqrt{u_m}}\Bigg].
  \end{split}
\end{equation}
If it arrives to the left edge $u=0$ of the dispersionless solution at the moment
\begin{equation}\label{t5-60.13}
   t_2(0)=\frac{l}{12u_mk_0}  \ln\frac{k_0+2\sqrt{u_m}}{k_0-2\sqrt{u_m}}
\end{equation}
before the wave breaking time (\ref{t5-58.3b}), $t_2(0)\leq t_b$, that is if
$k_0\geq k_0^c$, where $k_0^c$ is the root of the equation
\begin{equation}\label{t5-60.14}
  \frac{\sqrt{u_m}}{k_0^c} \ln \frac{k_0^c+2\sqrt{u_m}}{k_0^c-2\sqrt{u_m}}= 1,\quad
  \text{or}\quad
   k_0^c\approx 2.399\,\sqrt{u_m},
\end{equation}
then the packet moves all the time along the smooth profile without getting into the
dispersive shock region. For the wave number in the interval $2\sqrt{u_m}<k_0<k_0^c$
the packet's path goes partly through the dispersive shock region and this part
of the path is outside of the applicability domain of our theory. In Fig.~\ref{fig7}
we show the results of the numerical solution of the KdV equation for the negative pulse
with $u_m=1$ and the initial carrier wave number $k_0=2.4$ which satisfies the condition
$k_0>k_0^c$ (see Eq.~(\ref{t5-60.14})), so that the packet considered as a point
particle must leave
the region of the smooth pulse before its wave breaking moment. In Fig.~\ref{fig7}(a)
the whole wave structure is depicted at three moments of time: the initial one
(red) with the packet at the right edge of the smooth pulse, at the moment $t=t_1(u_m)$
(blue) with the packet at the minimal point of the smooth pulse distribution; and
at the moment close to the exit time $t=t_2(0)$ (black) when the packet is partly
inside the smooth pulse and partly outside it. To see better the evolution of the
profile of the linear wave packet, its profile was found by subtraction of
the solution of the KdV equation without initial high frequency wave packet from
the total wave amplitude Fig.~\ref{fig7}(a) obtained numerically with account the initial
wave packet, and the results of such a procedure at the same moments of time are shown
in Fig.~\ref{fig7}(b). As we see, the wave packet widens with time due to the second
order dispersion effects which are outside the scope of our approach.
Although the wave numbers distribution widens, the evolution of its mean value agrees
with our theoretical predictions. As one can see in Fig.~\ref{fig8}, the spectrum
considerably deforms during the packet's motion, especially at the moment of time
when the wide packet spreads partly outside of the smooth pulse and hardly can be
approximated by a narrow point-like structure. Nevertheless, the spectral distribution
has a sharp peak at the value $k=k_0$, as it should be expected at the exit point of 
the packet from the smooth pulse region. Not to close to the entrance and exit points 
the spectrum is localized quite well and evolution of the central carrier wave number 
follows the theoretical predictions, as it is illustrated in Fig.~\ref{fig9}.

\begin{figure}[t]
\begin{center}
\begin{picture}(130.,120.)
\put(-55,0){\includegraphics[width = 4.1cm]{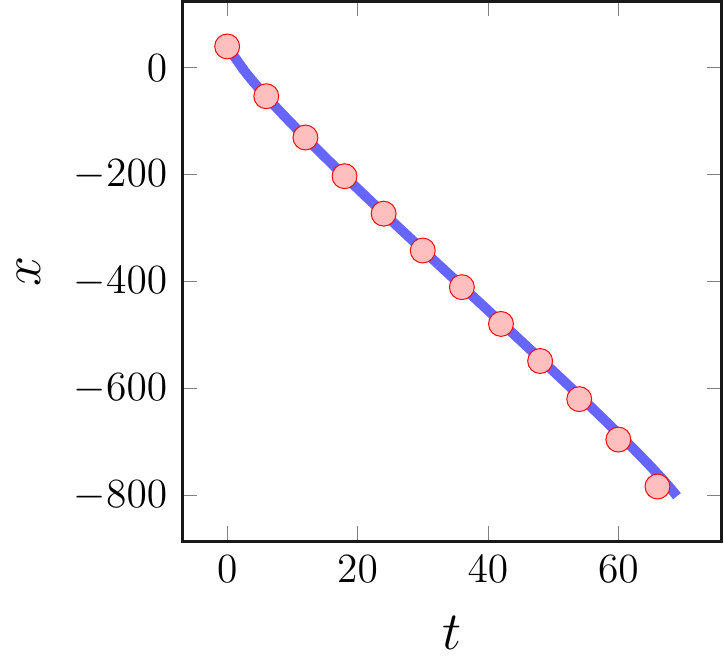}}
\put(65,0){\includegraphics[width = 4cm]{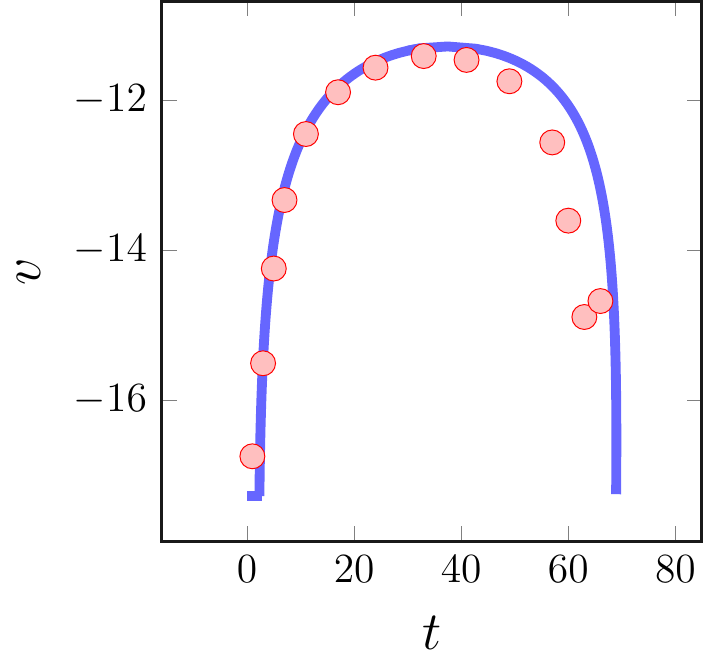}}
\put(5,115){(a)}
\put(130,115){(b)}
\end{picture}
\caption{
Coordinate (a) and velocity (b) of the wave packet propagating along evolving
negative pulse.
}
\label{fig10}
\end{center}
\end{figure}

Velocity of the packet as a function of time is determined by the parametric formulas
\begin{equation}\label{t5-60.15}
  v_g(u)=-6u(x(u),t_{1,2}(u))-3k_0^2,\qquad t=t_{1,2}(u),
\end{equation}
where $x(u)=6ut_{1,2}(u)+\ox_{1,2}(u)$, so we get $v_g(u)<-3k_0^2$.  Consequently,
interaction with smooth negative pulse decelerates the packet and it will be
behind a similar packet propagating with the velocity $v_g=-3k_0^2$ along a uniform
background with $u=0$ by the distance
\begin{equation}\label{t5-60.16}
  \Delta x_L=l\left(1-\frac{k_0}{4\sqrt{u_m}}
  \ln\frac{k_0+2\sqrt{u_m}}{k_0-2\sqrt{u_m}}\right).
\end{equation}
We call it again the ``phase shift'' caused by the packet-flow interaction.
The path of the wave packet and the corresponding velocity are compared with
the results of numerical simulations in Fig.~\ref{fig10}.

\subsection{Propagation of a wave packet along a positive pulse}

\begin{figure}[t]
\begin{center}
\includegraphics[width=7cm]{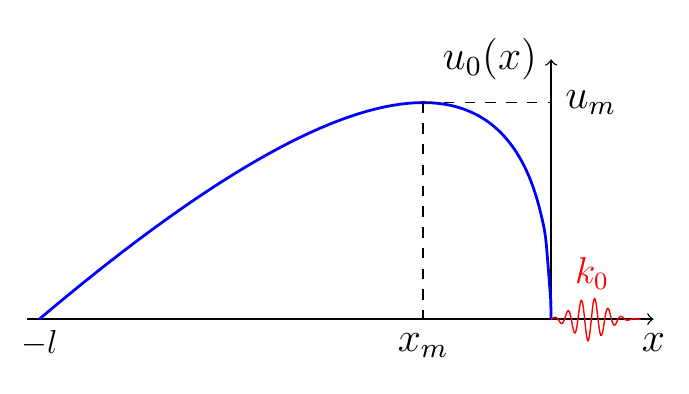}
\caption{Thick (blue) line shows the initial positive profile (\ref{t5-61.17}).
The wavy (red) line symbolizes a short wavelength wave
packet with the carrier wave number $k_0$ entering into the region of the
large scale smooth wave from the right.
}
\label{fig11}
\end{center}
\end{figure}

The theory of propagation of a high-frequency wave packet along a positive pulse
is somewhat simpler. If we take the initial background profile in the form shown in
Fig~\ref{fig11}, then it breaks at once at $t=0$, but the packet with the wave
number $k_0$ propagates to the left faster than the left edge of the appearing here
dispersive shock wave, so the packet considered as a point-like structure moves
along a smooth profile. Now for illustration
of the theory we take the initial distribution in the form
\begin{equation}\label{t5-61.17}
  u_0(x)=4u_m\left(\frac{x}l+\sqrt{-\frac{x}l}\right),\qquad -l\leq x\leq 0,
\end{equation}
so that
\begin{equation}\label{t5-61.18}
\begin{split}
  &\ox_1(u)=\frac{l}4\left(\frac{u}{u_m}-2+2\sqrt{1-\frac{u}{u_m}}\right),\\
  &\ox_2(u)=\frac{l}4\left(\frac{u}{u_m}-2-2\sqrt{1-\frac{u}{u_m}}\right)
  \end{split}
\end{equation}
and
\begin{equation}\label{t5-61.19}
\begin{split}
  &\ox'_1(u)=\frac{l}{4u_m}\left(1-\frac1{\sqrt{1-u/u_m}}\right),\\
  &\ox'_2(u)=\frac{l}{4u_m}\left(1+\frac1{\sqrt{1-u/u_m}}\right).
  \end{split}
\end{equation}
The dispersionless solution $x-6ut=\ox(u)$ can be transformed to the explicit
formula for $u(x,t)$:
\begin{equation}\label{t5-61.20}
\begin{split}
  u(x,t)=&\frac{4u_m}{(l+24u_mt)^2}\Big[12u_mt(l+2x)+lx+\\
  &+l\sqrt{24u_mt(6u_mt-x)-lx}\Big],
  \end{split}
\end{equation}
and the packet's motion along its two branches (\ref{t5-61.18}) is described by the
solutions of Eq.~(\ref{t5-59.5}) with the right-hand sides given by Eqs.~(\ref{t5-61.19}).
As a result we obtain the expressions for the moments of time at which the packet
reaches the points with the amplitude $u$ of the evolving smooth background along
each branch of the solution:
\begin{equation}\label{t5-61.21}
\begin{split}
   &t_1(u)=\frac{l}{24u_m\sqrt{k_0^2+4u}}\Bigg\{k_0-\sqrt{k_0^2+4u}+\\
  &+2\arcsin\sqrt{\frac{k_0^2+4u}{k_0^2+4u_m}}-2\arcsin\frac{k_0}{\sqrt{k_0^2+4u_m}}\Bigg\},\\
    &t_2(u)=\frac{l}{24u_m\sqrt{k_0^2+4u}}\Bigg\{k_0-\sqrt{k_0^2+4u}+2\pi-\\
  &-2\arcsin\sqrt{\frac{k_0^2+4u}{k_0^2+4u_m}}+2\arcsin\frac{k_0}{\sqrt{k_0^2+4u_m}}\Bigg\}.
  \end{split}
\end{equation}
The formula $x(u)=6ut(u)+\ox(u)$ for the location of the packet together with
Eqs.~(\ref{t5-61.21}) determines the law of motion of the point-like packet.
The positive background accelerates the packet
and it reaches the opposite edge of the background distribution at the moment $t_2(0)$.
Hence, in this case the phase shift is positive and equal to
\begin{equation}\label{t5-61.22}
\begin{split}
  \Delta x_L&=l\left\{1-\frac{k_0}{4\sqrt{u_m}}\left(\pi-2\arctan\frac{k_0}{2\sqrt{u_m}}\right)\right\}\\
  &=l\left\{1-\frac{ik_0}{4\sqrt{u_m}}\ln\frac{ik_0+2\sqrt{u_m}}{ik_0-2\sqrt{u_m}}\right\}.
  \end{split}
\end{equation}
The last expression can be obtained from Eq.~(\ref{t5-60.16}) by means of the
replacement $k_0\to ik_0$.

\begin{figure}[t]
\begin{center}
\begin{picture}(130.,120.)
\put(-58,0){\includegraphics[width = 3.95cm]{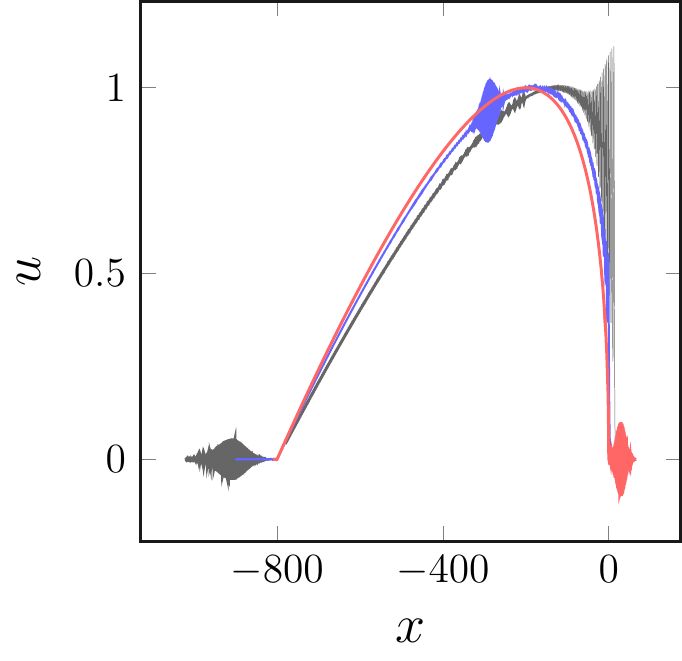}}
\put(62,0){\includegraphics[width = 4.3cm]{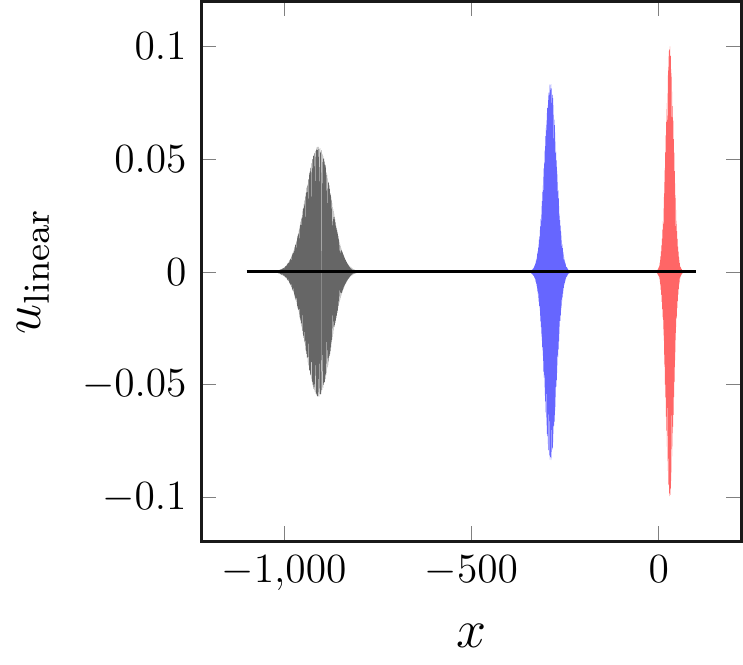}}
\put(5,115){(a)}
\put(130,115){(b)}
\end{picture}
\caption{
(a) The total numerical profiles $u(x,t)$ obtained by numerical solution of
the KdV equation for $u_m=1$, $l=800$, $k_0=5$ at the moments of time $t=0$ (red),
$t=4$ (blue), $t=12$ (black). Oscillatory region at the right edge of the wave
structure for $t=12$ corresponds to the dispersive shock wave generated after the
wave breaking moment. (b) The profiles of the linear wave packets
obtained by subtraction of the smooth evolution of the background pulse at
the same moments of time.
}
\label{fig12}
\end{center}
\end{figure}

\begin{figure}[t]
\begin{center}
\includegraphics[width=5cm]{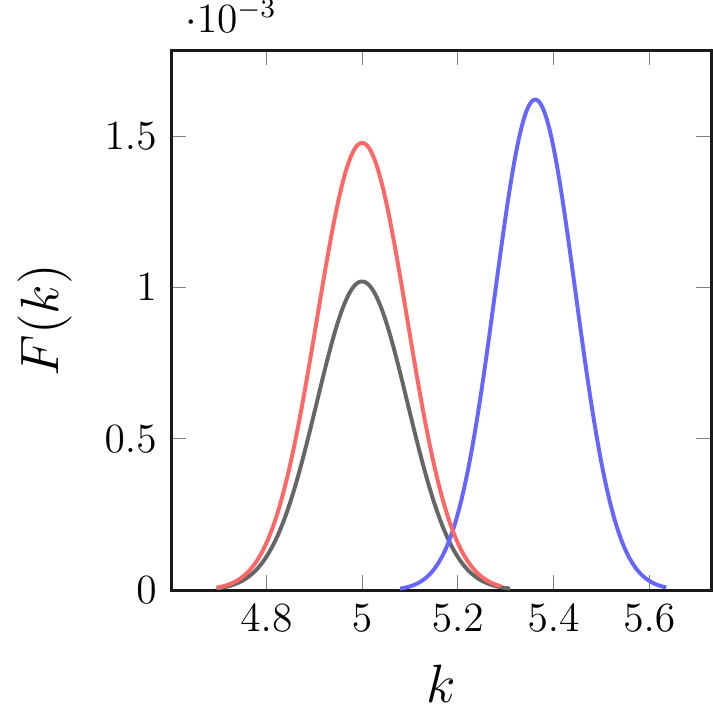}
\caption{
The numerically obtained spectrum (\ref{spectrum}) of the linear packet for
the three moment of time $t=0$ (red), $t=4$ (blue), $t=12$ (black). As we see, at 
the moment $t=12$ when the packet leaves the non-uniform smooth pulse region, the 
carrier wave number returns to its initial value $k_0=5$ in agreement with the theory.
}
\label{fig13}
\end{center}
\end{figure}

\begin{figure}[t]
\begin{center}
\includegraphics[width=5cm]{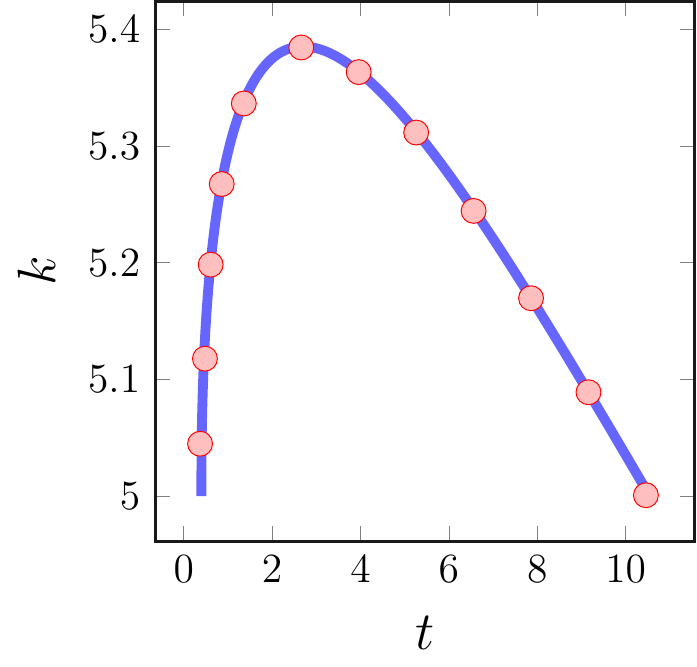}
\caption{
The dependence of the carrier wave number on time for the
positive smooth pulse and the initial wave number $k_0=5$.
}
\label{fig13}
\end{center}
\end{figure}

\begin{figure}[t]
\begin{center}
\begin{picture}(130.,120.)
\put(-55,0){\includegraphics[width = 4.12cm]{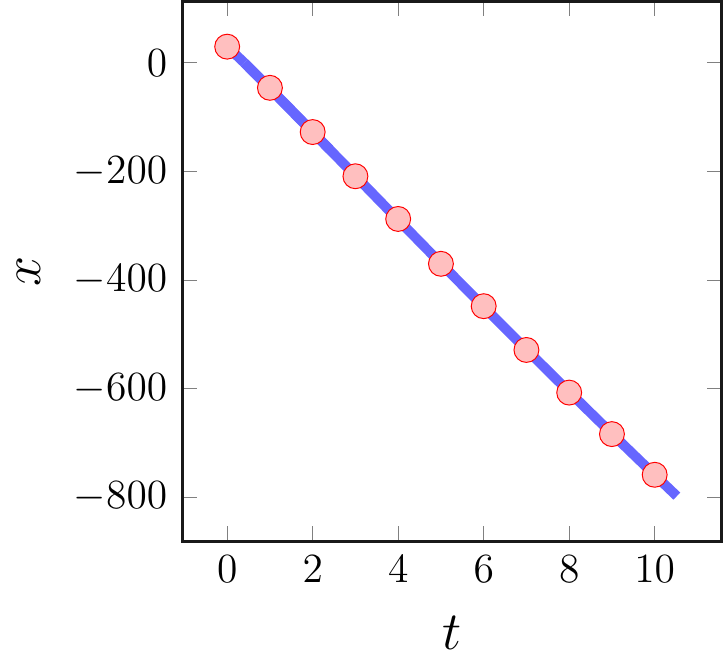}}
\put(65,0){\includegraphics[width = 4cm]{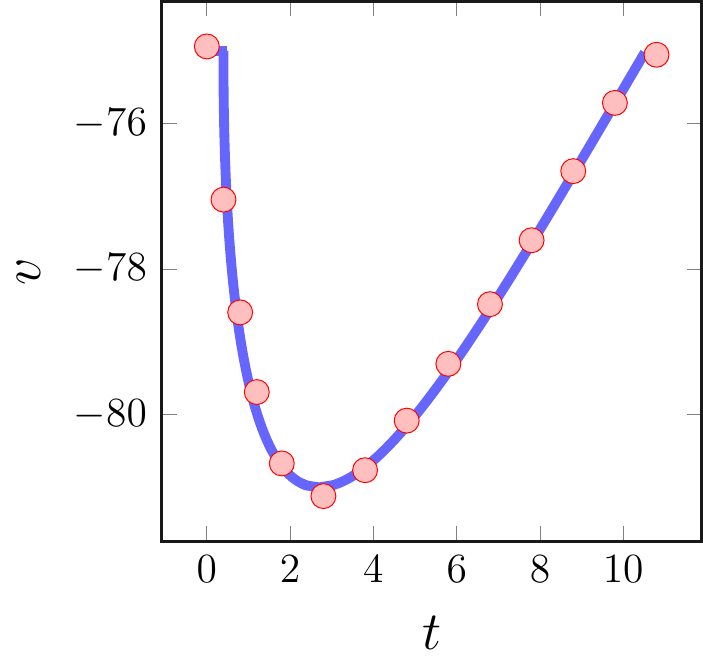}}
\put(5,115){(a)}
\put(130,115){(b)}
\end{picture}
\caption{
Coordinate (a) and velocity (b) of the wave packet propagating along evolving
positive pulse.
}
\label{fig14}
\end{center}
\end{figure}

Interaction of the wave packet with the positive smooth pulse is qualitatively the same for any
large enough value of $k_0$. However, in comparison with numerical results we often have to take
into account a finite size of the packet and therefore it is necessary to make the following
reservation related with a specific form of the initial distribution (\ref{t5-61.17}) which
breaks at the same moment of time $t=0$ as the packet starts its motion according to
Fig.~\ref{fig11}. Consequently, at the initial stage of propagation the packet with $k_0\sim1$
overlaps due to its finite size with the dispersive shock wave. This leads to noticeable
deformation of the spectrum of the packet which
introduces some uncertainty in numerical evaluation of the carrier wave number. To avoid such
inessential complications, we have chosen in our numerics a large enough initial wave number
$k_0=5$, so that the duration of interaction of the packet with the dispersive shock wave
becomes very small. We show in Fig.~\ref{fig12} the full amplitude $u(x,t)$ for three moments
of time $t=0,\, 4,\,12$ (a) and the linear packet's amplitude for
the same moments of time (b) obtained by means of the subtraction procedure described above
for the propagation of the packet along the negative background pulse. Although the packet widens
during its propagation, the spectrum changes very little and it remains being localized around
the initial value $k_0=5$ after the packet's exit from the smooth pulse region: see
Fig.~\ref{fig13}.
The packet's coordinate and its velocity agree
very well with theoretical predictions; see Fig.~\ref{fig14}.

\section{Conclusion}

In this paper, we have developed the theory of propagation of short wavelength packets along the
evolving smooth background under supposition that this evolution belongs to the simple-wave type
of dispersionless flows. Large difference of wavelength scales in the packet and in the
smooth pulse allows one to combine the Hamilton equations describing propagation of packets
in the geometrical optics approximation with the Hopf equation describing the dispersionless
evolution of the simple wave. As an immediate consequence of this combined system, we obtain
Eq.~(\ref{t5-56.4}) derived earlier by G.~A.~El in Ref.~\cite{el-05} from the small-amplitude
limit of Whitham equations which describe evolution of dispersive shock waves in the
Gurevich-Pitaevskii approach to the dispersive shocks theory developed in Ref.~\cite{gp-1973} (see also
Ref.~\cite{kamch-21b} and references therein). Solution of this equation with the initial condition
$k=k_0$ at $u=u_0$ yields the dependence $k=k(u)$ of the carrier wave number $k$ on the value
$u$ of the background flow amplitude at the instant location of the packet. Then, as was shown
in Ref.~\cite{kamch-19a}, the path of the packet can be found from the condition that the
packet's motion with the group velocity must be consistent with the known solution of
the Hopf equation for the background flow. This method was applied earlier to description
of motion of small-amplitude edges of dispersive shock waves in wave systems obeying
non-integrable equations (the shallow water Serre equation in Ref.~\cite{ik-19} and
equations for ion-sound nonlinear plasma waves in Ref.~\cite{ik-20}), and here we extend
it arbitrary localized wave packets.

The developed in this paper approach is illustrated by its application to wave systems
whose evolution is described by the KdV equation. This type of problems has been recently
discussed in Ref.~\cite{ceh-19} for a particular case of a rarefaction wave background
flow. We have extended the theory to the general form of simple wave flows. The analytical
theory was illustrated by an examples of the initial distributions which admit
simple and explicit analytical solutions. Comparison
of our analytical theory with the results of numerical solutions of the KdV equation
with appropriate initial conditions demonstrates quite good accuracy of our approximation.
One may suppose that this approach will find application to various problems of wave
packets propagation along evolving smooth pulses of a simple wave type.

\begin{acknowledgments}
We thank S.~K.~Ivanov and Y.~A.~Stepanyants for useful discussions and important remarks.
The reported study was funded by RFBR, project number 20-01-00063.
\end{acknowledgments}

\end{document}